\documentclass[12pt,a4paper]{article}
\begin{document}
\parindent=0pt
\parskip=3mm

\renewcommand{\baselinestretch}{1.0}

\def\ref#1{\noindent \parshape=2 0pt 442pt 16pt 426pt #1}
\def\reef#1{\noindent \parshape=2 -8pt 530pt 16pt 506pt #1}
\def\rof#1{\noindent \parshape=2 20pt 502pt 35pt 487pt #1}
\def\rif#1{\noindent \parshape=2 16pt 486pt 16pt 486pt #1}
\def\ruf#1{\noindent \parshape=2 16pt 490pt 16pt 490pt #1}

$\,$

\vskip 2.0cm

\begin{center}
\LARGE
{\bf Azimuthally symmetric MHD and two--fluid equilibria with arbitrary flows}
\\[2ex]

\Large 

\vskip 2.0cm

K. G. McClements\footnote{k.g.mcclements@ukaea.org.uk}
and A. Thyagaraja\footnote{a.thyagaraja@ukaea.org.uk}  
\\[2ex]

\vskip 2.0cm

EURATOM/UKAEA Fusion Association, Culham Science Centre, Abingdon, OX14 3DB, 
UK\\

\vskip 2.0cm

\normalsize

Submitted to {\it Monthly Notices of the Royal Astronomical Society}\\

UKAEA Fusion Report UKAEA FUS 430\\

\end{center}
\newpage

\begin{center}
\Large
{\bf Abstract}
\end{center}
\normalsize

Magnetohydrodynamic (MHD) and two--fluid quasi-neutral equilibria with 
azimuthal symmetry, gravity and arbitrary ratios of (nonrelativistic) flow 
speed to acoustic and Alfv\'en speeds are investigated. In the two--fluid 
case, the mass ratio of the two species is arbitrary, and the analysis is 
therefore applicable to electron--positron plasmas. The methods of derivation 
can be extended in an obvious manner to several charged species. Generalized 
Grad--Shafranov equations, describing the equilibrium magnetic field, are 
derived. Flux function equations and Bernoulli relations
for each species, together with Poisson's equation for the gravitational
potential, complete the set of equations required to determine the 
equilibrium. These are straightforward to solve numerically. The two--fluid 
system, unlike the MHD system, is shown to be free of singularities. It is 
demonstrated analytically that there exists a class of incompressible MHD 
equilibria with magnetic field--aligned flow. A special sub--class 
first identified by S. Chandrasekhar, in which the flow speed is
everywhere equal to the local Alfv\'en speed, is compatible with virtually any
azimuthally symmetric magnetic configuration. Potential applications of this 
analysis include extragalactic and stellar jets, accretion disks, and plasma 
structures associated with active late--type stars.

\vskip 0.5cm


\newpage
\normalsize

\section{Introduction}

Observations across a range of wavelengths dating back over several decades 
have established the existence of jets associated with stars (Camenzind 
1998; Henriksen 1998) and active galactic nuclei (Zensus 1997; Ferrari 1998). 
These structures, and also accretion disks, have in common the 
following features: they contain flows of bulk matter; they are embedded in 
magnetic fields; and they can be treated, to a first approximation, as having 
an axis of symmetry. All three features are also found in most of the 
laboratory magnetic confinement systems which have been investigated 
experimentally, such as tokamaks (Wesson 1997). There is increasing evidence 
generally that magnetic fields play an important role in governing 
the physics of a wide variety of astrophysical processes. These include, for 
example, the formation and early evolution of galaxies, and interstellar gas
dynamics (Zweibel \& Heiles 1997). In the case of accretion disks, it 
is believed that purely hydrodynamic models cannot provide the effective 
viscosity required to account for observationally--inferred accretion rates, 
and that magnetohydrodynamic (MHD) effects must be taken into account (Hawley 
\& Stone 1998). Magnetic fields may also provide a means of connecting 
accretion disks with jets, via the generation of MHD waves (Tagger \& Pellat 
1999). Finally, evidence has emerged recently that plasma confinement by 
dipole--like magnetic fields can account for X--ray and radio emission from 
active late--type stars (Kellet, Bingham \& Tsikoudi 2000).  

The first step in the construction of a theoretical model of a quasi--steady
magnetized plasma structure (whether astrophysical or in the laboratory) is a 
determination of its equilibrium state. In general, this requires all time 
derivatives to be set equal to zero in the combined system of Maxwell and
magnetized fluid equations, and solutions determined for the magnetic field, 
density, and, if applicable, flow velocity in three dimensions. Having 
determined the equilibrium configuration, observed time variations can then 
often be interpreted theoretically as perturbations of the equilibrium state.  
MHD equilibrium studies of axisymmetric systems with flow have been carried 
out by many authors, including Chandrasekhar (1956), Woltjer (1959), Morozov 
and Solov'ev (1963), Zehrfeld \& Green (1972), Maschke \& Perrin (1980)
and Throumoulopoulos \& Pantis (1989). Early work in this field 
was restricted to the case of incompressible flow (Chandrasekhar 1956; 
Woltjer 1959). The problem of MHD equilibrium in toroidal systems with 
compressible flow was studied in the ideal limit by Morozov \& Solov'ev (1963),
and in the resistive case by Zehrfeld \& Green (1972). The case of purely
toroidal flow has been studied by Maschke \& Perrin (1980) and 
Throumoulopoulos \& Pantis (1989). Self--similar flow solutions were obtained 
by Blandford \& Payne (1982), and Lovelace et al. (1986) developed a general 
theory of axisymmetric MHD equilibria with relativistic flows, which they 
applied to disks associated with rotating magnetized stars and black holes. 
Rosso \& Pelletier (1994) used a variational method to resolve mathematical
problems arising from MHD flow singularities. Bogoyavlenskij (2000) recently 
demonstrated the existence of exact axisymmetric MHD equilibria which do not 
include flow, but may nevertheless have applications to astrophysical jets. 
Krasheninnikov et al. (2000) included flow effects in a study of magnetic 
dipole equilibria, while Keppens \& Goedbloed (2000) examined axisymmetric
stellar wind equilibria with both open and closed magnetic field regions. 

These analyses of axisymmetric MHD equilibria are generally based on 
the Grad--Shafranov equation, derived originally to describe magnetic
field equilibria in nuclear fusion experiments (Shafranov 1958; Grad \& Rubin
1958): it is essentially an expression of momentum balance for one or more 
magnetized fluids. The analysis in this paper is also based on various forms 
of the Grad--Shafranov equation: we investigate ideal axisymmetric MHD 
(Section 2) and two--fluid (Section 3) equilibria with flows which are 
nonrelativistic but otherwise arbitrary. Whereas MHD equilibria have been 
studied in considerable detail by previous authors (in particular, Lovelace et
al. 1986), little attention has been paid to two--fluid effects. We use a 
similar formalism for the MHD and two--fluid models, thus making it 
straightforward (and instructive) to compare and contrast them. The 
two--fluid model provides the basis for a more comprehensive description of 
jets and accretion disks than the MHD model, and is actually more tractable 
numerically. In the MHD case, special classes of solutions can be identified 
analytically (Section 4), which provide useful benchmarks for more realistic 
numerical solutions. 

\section{General equilibrium analysis: MHD}

We present an alternative derivation of the ``generalised Grad--Shafranov'' 
equation of ideal MHD with arbitrary flows in azimuthally symmetric systems. 
This equation was first obtained by Lovelace et al. (1986): our alternative
derivation is simpler than that of Lovelace et al., and can be readily 
generalized to the two--fluid case. A similar analysis was carried out by 
Goedbloed \& Lifschitz (1997) for a system with translational rather than
azimuthal symmetry. We consider non--relativistic MHD 
equilibria of a quasi--neutral plasma. The crucial simplifications are due to 
the assumed azimuthal symmetry (about the $z$--axis) and steady conditions. We
adopt a cylindrical coordinate system, denoting the azimuthal angle by $\phi$, 
and distance from the symmetry axis by 
$r$. It is useful to consider an ``external'' source of gravitation, creating 
an azimuthally symmetric gravitational potential $V(r,z)$, although it will be
seen that Newtonian self--gravitation of the plasma can be easily 
incorporated into the analysis.   

The Maxwell equation $\nabla\cdot{\bf B}=0$ ensures that the magnetic field 
${\bf B}$ has potential representation
$${\bf B} = \left[ -\frac{1}{r}\frac{\partial \Psi}{\partial z} {\bf
  e}_{r}+B_{\phi}{\bf e}_{\phi}+\frac{1}{r}\frac{\partial \Psi}{\partial r}
{\bf e}_{z} \right]. \eqno (1) $$
In equation (1) $\Psi(r,z)$ is the poloidal magnetic flux function and 
$B_{\phi}(r,z)$ is the toroidal field. It is easily shown that $\Psi/r$ is 
the toroidal (azimuthal) component of the magnetic vector potential {\bf A}. 
Since ${\bf B}\cdot \nabla\Psi = 0$, the function $\Psi$ is constant along 
magnetic field lines.  
In an analogous manner, mass continuity ensures that the plasma mass flux 
vector, defined to be the product of mass density $\rho$ and flow velocity
{\bf v}, can be represented in the form
$$ \rho {\bf v} = \left[ -\frac{1}{r}\frac{\partial \chi}{\partial z} {\bf
e}_{r}+\rho v_{\phi}{\bf e}_{\phi}+\frac{1}{r}\frac{\partial \chi}{\partial r}
{\bf e}_{z} \right]. \eqno (2) $$
where $\chi$ is the mass flow function. Because of the assumed azimuthal
symmetry, this is simply related to the poloidal magnetic flux function $\Psi$
as follows. The ideal MHD Ohm's law in Gaussian cgs units is
$$ \nabla\Phi = \frac{{\bf v}}{c}\times {\bf B}, \eqno (3) $$
where $c$ is the speed of light and $\Phi$ has gradient equal to minus the 
electric field {\bf E}: the 
existence of such a potential follows from the steady--state assumption. 
Since $\Phi$ cannot depend on $\phi$, the azimuthal component of equation (3) 
yields
$$ v_rB_z = v_zB_r. $$
Expressing the velocity and magnetic field components in terms of $\Psi$ and 
$\chi$ using equations (1) and (2), we obtain
$$\frac{\partial(\chi,\Psi)}{\partial(r,z)} = 0, $$
from which it follows that 
$$\chi = F(\Psi), \eqno (4) $$
where $F$ is an arbitrary function. This, and other arbitrary functions 
appearing in the MHD and two--fluid systems of equations, are determined 
ultimately by plasma transport processes (Freidberg 1982). However, 
valuable physical insights can often be gained by adopting simple forms for 
these functions (see, e.g., Bogoyavlenskij 2000).     

The $r$ and $z$ components of equation (3) can be written in the form
$$ -c\frac{\partial \Phi}{\partial r} = \frac{(\frac{B_{\phi}F^{\prime}}
{\rho}-v_{\phi})}{r}\frac{\partial\Psi}{\partial r}, \eqno (5) $$
and
$$ -c\frac{\partial \Phi}{\partial z} = \frac{(\frac{B_{\phi}F^{\prime}}
{\rho}-v_{\phi})}{r}\frac{\partial\Psi}{\partial z}, \eqno (6) $$ 
where $F^{\prime} \equiv dF/d\Psi$. Introducing the quantity 
$\Omega \equiv (v_{\phi}-B_{\phi}F^{\prime}/\rho)/r$, and
eliminating $\Phi$ in the two equations above by cross--differentiation and
subtraction, we find that
$$ \frac{\partial(\Omega,\Psi)}{\partial(r,z)} = 0. $$
This indicates that $\Omega$ is a function of $\Psi$. From the definition of 
$\Omega$, we thus have
$$ rv_{\phi}-{rB_{\phi}F^{\prime}\over \rho} = r^2\Omega(\Psi). \eqno (7) $$  
In a similar fashion,
by eliminating $\Omega$ from equations (5) and (6), it is straightforward to
show that $\Phi$ is also a function of $\Psi$ and that $c\Phi^{\prime}=\Omega$ 
(the prime again denoting differentiation with respect to $\Psi$).
These results have several interesting physical interpretations. First, 
azimuthally symmetric, steady MHD equilibria with flows have electrostatic 
potentials which are functions of the poloidal magnetic flux, regardless of 
centrifugal and Coriolis effects at arbitrary flow Mach number. 
Second, in any poloidal plane (i.e. any $(r,z)$ plane
with $\phi$ fixed), although not necessarily in three dimensional
space, the flow and magnetic field components are
parallel. Third, the poloidal mass flow function depends only on the poloidal
magnetic flux. However, in general the density is not a flux function.

To proceed further, it is necessary to calculate the components of the 
vorticity ${\bf K} \equiv \nabla \times {\bf v}$, the current density 
${\bf j}=(c/4\pi)\nabla \times {\bf B}$, and the vector products 
$(\nabla\times{\bf B})\times {\bf B}$, ${\bf K} \times {\bf v}$. The following 
relations are easily derived from the definitions:
$$ \nabla\times{\bf B} =  \left[ -\frac{\partial B_{\phi}}{\partial z}{\bf
e}_{r}+j^{*}_{\phi}{\bf e}_{\phi}+\frac{1}{r}\frac{\partial}{\partial r}
(rB_{\phi}){\bf e}_{z} \right], \eqno (8) $$
$$j^{*}_{\phi} =  -\frac{1}{r}\left[\frac{\partial^{2} \Psi}{\partial 
z^2}+r\frac{\partial}{\partial r}(\frac{1}{r}\frac{\partial
\Psi}{\partial r})\right], \eqno (9) $$
$${\bf K} = \left[ -\frac{\partial v_{\phi}}{\partial z}{\bf
e}_{r}+K^{*}_{\phi}{\bf e}_{\phi}+\frac{1}{r}\frac{\partial}{\partial r}
(rv_{\phi}){\bf e}_{z} \right], \eqno (10) $$
$$K^{*}_{\phi} = -\frac{1}{r}\left[\frac{\partial}{\partial
z}(\frac{F^{\prime}}{\rho}\frac{\partial \Psi}{\partial
z})+r\frac{\partial}{\partial r}(\frac{F^{\prime}}{r\rho}
\frac{\partial \Psi }{\partial r})\right], \eqno (11) $$ 
$$\leftline{$\displaystyle (\nabla \times {\bf B})\times {\bf B}
= \Biggl[(j^{*}_{\phi}\frac{1}{r}\frac{\partial \Psi}{\partial r}-
\frac{rB_{\phi}}{r^{2}}\frac{\partial}{\partial r}(rB_{\phi})){\bf e}_{r}
+\frac{1}{r^{2}}\frac{\partial (\Psi,rB_{\phi})}{\partial 
(r,z)}{\bf e}_{\phi}$}$$
$$\;\;\;\;\;\;\;\;\;\;\;\;\;\;\;\;\;\;\;\;\;\;\;
+\Bigl(j^{*}_{\phi}\frac{1}{r}\frac{\partial \Psi}{\partial z}-
\frac{rB_{\phi}}{r^{2}}\frac{\partial}{\partial z}(rB_{\phi})\Bigr)
{\bf e}_{z}\Biggr], \eqno (12) $$ 
$$\leftline{$\displaystyle {\bf K}\times {\bf v}
= \Biggl[(K^{*}_{\phi}\frac{F^{\prime}}{\rho r}\frac{\partial
\Psi}{\partial r}-\frac{rv_{\phi}}{r^{2}}
\frac{\partial}{\partial r}(rv_{\phi})){\bf e}_{r}+\frac{F^{\prime}}
{\rho r^{2}}\frac{\partial (\Psi,rv_{\phi})}{\partial (r,z)}{\bf e}_{\phi}$}$$
$$\;\;\;\;\;\;\;\;\;\;\;\;\;\;\;\;\;\;\;\;\;\;\;
+\Bigl(K^{*}_{\phi}\frac{F^{\prime}}{\rho r}\frac{\partial \Psi}{\partial z}
-\frac{rv_{\phi}}{r^{2}}
\frac{\partial}{\partial z}(rv_{\phi})\Bigr){\bf e}_z\Biggr] . \eqno (13) $$
The quantity $j^{*}_{\phi} = 4\pi j_{\phi}/c$,
where $j_{\phi}$ is the azimuthal current density.

The system of MHD equations is completed by the ideal isentropic
equation and the equation of motion. For a perfect gas with ratio of specific
heats $\gamma$ the specific entropy $s$ is a function of $p/\rho^{\gamma} 
\equiv \sigma$. In the absence of dissipation, $\sigma$ must be conserved 
in the fluid frame:
$$ {\bf v}\cdot\nabla\sigma = 0 . $$ 
It is clear from equations (2) and (4) that this condition is equivalent to
$$\frac{\partial(\sigma,\Psi)}{\partial(r,z)} = 0. \eqno (14) $$ 
The steady--state equation of motion is the usual Eulerian one, with Lorentz, 
pressure and gravitational forces included. In terms of vorticity {\bf K} it 
can be written in the form 
$${\bf K}\times {\bf v} = -\frac{1}{\rho}\nabla p-\nabla \frac{{\bf v}^2}{2}-
\nabla V +\frac{1}{4\pi \rho} (\nabla \times {\bf B})\times {\bf B}. 
\eqno (15) $$
To obtain equation (15) we have used the vector identity
$$ {\bf v}\cdot\nabla {\bf v} = (\nabla \times {\bf v})\times {\bf v}+\nabla 
\frac{{\bf v}^{2}}{2}. \eqno (16) $$
Equation (14) has the immediate consequence that $\sigma$ (a measure of the 
specific entropy) must be an arbitrarily specifiable flux function. 

We next consider the azimuthal component of the equation of motion.
Noting the fact that all quantities are independent of $\phi$, 
due to the assumed azimuthal symmetry, we infer from
equations (12), (13) and (15) that
$$ \frac{F^{\prime}}{r^{2}}\frac{\partial (\Psi,rv_{\phi})}{\partial (r,z)}
= \frac{1}{4\pi r^{2}}\frac{\partial (\Psi,rB_{\phi})}{\partial (r,z)}. $$
Cancelling $r^2$ in the denominator on both sides and observing that 
$F^{\prime}$ depends only upon $\Psi$, we see that this relation is equivalent
to
$$ \frac{\partial (F^{\prime}rv_{\phi}-\frac{1}{4\pi}rB_{\phi},\Psi)}{\partial
(r,z)} = 0. \eqno (17) $$
This relation, an expression of the law of conservation of canonical angular 
momentum of the fluid about the symmetry axis, connects the toroidal flow and 
magnetic field components. Defining
$$ \Lambda = F^{\prime}rv_{\phi}-\frac{1}{4\pi}rB_{\phi}, \eqno (18) $$
it is clear from equation (17) that $\Lambda$ is a flux function.

It will be seen that the quantities defined by equations (7) and (18) play 
a key role in the reduction of the equations of motion. Making use of equations
(12) and (13), we find that equation (15) in the $(r,z)$ plane can be written 
in the form
$$\leftline{$\displaystyle \left[K^{*}_{\phi}F^{\prime}-\frac{1}{4\pi}
j^*_{\phi}\right]\frac{\nabla \Psi}{\rho r} = \frac{1}{2r^{2}}\nabla
\left[(rv_{\phi})^{2}\right]-\frac{1}{8\pi \rho r^{2}}\left[\nabla 
(rB_{\phi})^{2}\right]-\frac{1}{\rho}\nabla p $}$$
$$ -\nabla \frac{{\bf v}^{2}}{2}-\nabla V. \eqno (19) $$
If the gravitational potential $V$ and the ``structure functions'', 
$F(\Psi)$, $\Omega(\Psi)$, $\Lambda(\Psi)$ and $\sigma(\Psi)$ are prescribed,
equation (19) represents two partial differential equations for the 
determination of $\Psi$ and $\rho$ as functions of $r,z$, subject to suitable
boundary conditions. The explicit forms of these equations will now be 
derived in full generality.

We note that the left hand side of equation (19) is annihilated by taking the 
scalar product with the vector $\nabla \Psi \times{\bf e}_{\phi}$, which is 
always in a poloidal plane and tangential to the flux curves defined by 
constant values of $\Psi$. 
The right hand side is simply related to the tangential derivatives of various 
quantities, by virtue of a geometrical relation valid for any function 
$f(r,z)$:
$$ (\nabla \Psi \times {\bf e}_{\phi}).\nabla f = \frac{\partial (\Psi,f)}{
\partial (r,z)} = {\mid}\nabla \Psi {\mid} \frac{\partial f}{\partial l}, 
\eqno (20)$$
where $l$ denotes arc length along the flux line. Using this, we annihilate 
the left hand side of equation (19) and derive the relation
$$\frac{1}{2r^{2}}\frac{\partial \left[(rv_{\phi})^{2}\right]}{\partial l}
-\frac{1}{8\pi \rho r^{2}}\frac{\partial \left[
(rB_{\phi})^{2}\right]}{\partial l} = \frac{\partial}{\partial l} 
\left[\frac{\gamma}{\gamma-1}(\sigma\rho^{\gamma -1}) + \frac{{\bf
v}^2}{2}+V\right]. \eqno (21) $$
This equation represents a far--reaching generalization of the well--known 
Bernoulli equation of gas dynamics. A formal integral of it will now be 
obtained. 

Taking account of equation (18), we introduce the following representations of 
$rv_{\phi}$ and $rB_{\phi}$:
$$rv_{\phi} = J(\Psi)+\Theta(r,z), \eqno (22) $$
$$rB_{\phi} = I(\Psi)+4\pi F^{\prime}(\Psi)\Theta(r,z). \eqno (23) $$
Equation (18) can then be written in the form
$$\Lambda(\Psi) = JF^{\prime}-\frac{I}{4\pi}. \eqno (24) $$
In equations (22) and (23) $\Theta$ is a function to be determined and $I,J$ 
are arbitrary functions of $\Psi$ related to $\Lambda$ and $\Omega$. 
Substitution of these equations into equation (7) yields the relation
$$ (J+\Theta)-\frac{F^{\prime}}{\rho}(I+4\pi F^{\prime}\Theta)
= r^{2}\Omega(\Psi). \eqno (25) $$
This equation expresses $\Theta$ in terms of the flux functions $\Omega$, $I$ 
and $J$, and in terms of $\rho$ and $r^2$, which are not necessarily flux
functions. The quantity $\Theta$ measures the variation of $rv_{\phi}$ and 
$rB_{\phi}$ on flux surfaces. In terms of $J$ and $\Theta$, the two terms on 
the the left hand side of equation (21) are given by 
$$\frac{1}{2r^{2}}\frac{\partial \left[(rv_{\phi})^{2}\right]}{\partial l}
= \frac{1}{2r^{2}}\frac{\partial}{\partial l} \left[(J^{2}+2J\Theta
+\Theta^{2})\right] = \frac{1}{r^{2}}\left[(J
+\Theta)\right]\frac{\partial \Theta}{\partial l} , $$
$$\frac{1}{8\pi\rho r^2}\frac{\partial 
\left[(rB_{\phi})^{2}\right]}{\partial l} = \frac{1}{8 \pi \rho r^{2}}
\frac{\partial}{\partial l} 
\left[(I^{2}+2I 4 \pi F^{\prime}\Theta +16 \pi^{2} (F^{\prime})^{2}\Theta^{2})
\right] $$
$$\;\;\;\; = \frac{1}{\rho r^{2}}\left[IF^{\prime}+4\pi (F^{\prime})^{2}
\Theta\right]\frac{\partial \Theta}{\partial l}. $$
Thus we have
$$\frac{1}{2r^{2}}\frac{\partial \left[(rv_{\phi})^{2}\right]}{\partial l}
-\frac{1}{8\pi \rho r^{2}}\frac{\partial \left[(rB_{\phi})^{2}\right]}
{\partial l} = \frac{1}{r^{2}}\left[ (J+\Theta)-\frac{F^{\prime}}{\rho}(I
+4\pi F^{\prime}\Theta)\right]\frac{\partial \Theta}{\partial l}
= \frac{\partial }{\partial l}(\Omega\Theta), \eqno (26) $$
where we have made use of equation (25) and the fact that $\Omega$ is a flux
function (i.e. $\partial\Omega/\partial l=0$). Equation (21) can now be
exactly integrated by writing it in the form
$$\frac{\partial}{\partial l}\left[ \Omega \Theta-\frac{\gamma}{\gamma -1}
(\sigma\rho^{\gamma -1}) - \frac{{\bf v}^{2}}{2}-V\right] = 0. \eqno (27) $$ 
From equation (25) it follows that $\Theta$ is given by
$$ \Theta = \frac{r^{2}\Omega+(\frac{F^{\prime}}{\rho}I-J)}{1-\frac{4\pi 
(F^{\prime})^{2}}{\rho}}. \eqno (28) $$
It is clear from equation (27) that there exists a generalized Bernoulli 
integral 
$$\Omega \Theta-\frac{\gamma}{\gamma -1}(\sigma\rho^{\gamma -1}) - \frac{{\bf
v}^2}{2}-V \equiv -h(\Psi), \eqno (29) $$
where the arbitrary function $h(\Psi)$ may be regarded as an MHD 
generalization of the ``stagnation enthalpy'' (total pressure) of gas 
dynamics (Meyer 1971). In the present calculation it is convenient to use
the flux function
$$ H(\Psi) \equiv h(\Psi)-\Omega(\Psi)J(\Psi), \eqno (30) $$    
rather than $h$ itself. The Bernoulli relation then becomes
$$\frac{{\bf v}^2}{2}+V = H(\Psi)+\Omega rv_{\phi}-{\gamma\over\gamma-1}
\sigma\rho^{\gamma-1}. \eqno (31) $$

Using the Bernoulli relation in the form given by equation (31) to 
eliminate ${\bf v}^2/2+V$, we find that equation (19) can be written in the 
form 
$$\left[K^{*}_{\phi}F^{\prime}-\frac{1}{4\pi}j^*_{\phi}-
{(rB_{\phi})\over r}\Lambda^{\prime}+{(rB_{\phi})(rv_{\phi})\over r}F^{\prime
\prime}+{\rho r}H^{\prime}+\rho r(rv_{\phi})\Omega^{\prime}-{rp\over \gamma-1}
{\sigma^{\prime}\over \sigma}\right]\nabla \Psi
= 0. \eqno (32) $$
Nontrivial solutions of this equation have $\nabla\Psi\ne 0$: setting the 
bracketed quantity in equation (32) equal to zero, and using our expressions
for $K_{\phi}^*$, $j_{\phi}^*$ [equations (9) and (11)], we infer that    
$\Psi$ and $\rho$ satisfy 
$$\leftline{$\displaystyle \frac{\partial^{2} \Psi}{\partial z^{2}}+
r\frac{\partial}{\partial r}
(\frac{1}{r}\frac{\partial\Psi}{\partial r})-4\pi F^{\prime}
\left[\frac{\partial}{\partial z}(\frac{F^{\prime}}{\rho}\frac{\partial 
\Psi}{\partial z})+r\frac{\partial}{\partial r}(\frac{F^{\prime}}{r\rho}
\frac{\partial \Psi}{\partial r})\right]$}$$
$$+4\pi\rho r^2H^{\prime}+4\pi \rho r^2
\Omega^{\prime}rv_{\phi}-{4\pi r^2\over \gamma-1}\sigma^{\prime}
\rho^{\gamma}-4\pi rB_{\phi}\Lambda^{\prime}+4\pi(rB_{\phi})(rv_{\phi})
F^{\prime\prime} = 0. $$
This is a generalized form of the Grad-Shafranov equation. It is convenient to
rearrange it in the form
$$\leftline{$\displaystyle \frac{\partial}{\partial z}(\Delta \frac{\partial
\Psi}{\partial z})+r\frac{\partial}{\partial r}(\Delta \frac{1}{r}
\frac{\partial\Psi}{\partial r})+4\pi F^{\prime}F^{\prime
\prime}\frac{(\Psi_{r}^{2}+\Psi_{z}^{2})}{\rho} $}$$
$$+4\pi\rho r^2H^{\prime}+4\pi\rho r^2
\Omega^{\prime}rv_{\phi}-{4\pi r^2\over \gamma-1}\sigma^{\prime}
\rho^{\gamma}-4\pi rB_{\phi}\Lambda^{\prime}+4\pi(rB_{\phi})(rv_{\phi})
F^{\prime\prime} = 0, \eqno (33) $$ 
where
$$ \Delta(\rho,F^{\prime}) = 1-\frac{4\pi (F^{\prime})^{2}}{\rho}. 
\eqno (34) $$
It is straightforward to verify that Eq. (33) is identical to the  
Grad--Shafranov equation obtained by Lovelace et al. (1986).  
Given the arbitrary flux functions $F$, $\Lambda$, $\Omega$, $H$, $\sigma$ and
the potential $V$, equations (7), (18), (31), (33) and (34) determine $\Psi$ 
and $\rho$. The coefficients of the highest order derivatives in 
equation (33) vanish, and hence the equation becomes singular, when $\Delta = 
0$: physically, this corresponds to the projection of the flow velocity onto 
any $(r,z)$ surface being equal to an Alfv\'en speed defined in terms of the 
$(r,z)$ components of {\bf B}. The fact that the density $\rho$ depends, via 
the Bernoulli relation [equation (31)], on $\Psi$ and its derivatives means 
that equation (33) contains other singularities (Lovelace et al. 1986), which 
are less immediately apparent than the one corresponding to $\Delta = 0$.
The presence of $\rho$ in equation (33) also 
means that it is fundamentally nonlinear, regardless of the choice of 
arbitrary functions: in this respect it differs from the Grad--Shafranov 
equation without flows (e.g. Bogoyavlenskiy 2000). 
The existence of flow singularities aggravates considerably the 
difficulties involved in finding solutions (see e.g. Rosso \& Pelletier 1994).
In principle, the singularities can be removed by invoking dissipation or 
electron inertia, or by special, compatible choices of the arbitrary functions
involved.  

We have so far assumed that the gravitational potential $V$ is due to a
distribution of masses which is external to the plasma. It is straightforward 
to include Newtonian self--gravitation of the plasma by adding the 
gravitational Poisson equation
$$\frac{1}{r}\frac{\partial}{\partial r}(r\frac{\partial V}{\partial
r})+\frac{\partial^{2} V}{\partial z^{2}} = 4\pi G(\rho +\rho_{\rm ext}),
\eqno (35) $$ 
where $\rho_{\rm ext}$ represents the density distribution of external,
uncharged bodies (e.g. a neutron star or black hole). The Poisson integral
solution of this equation can be used to eliminate $V$ from the generalized 
Bernoulli relation [equation (31)]: this would have the effect of making 
the governing equations non--local.

\section{General equilibrium analysis: two--fluid theory}         

It is possible to generalize the above model to include two--fluid effects.
There are both physical and mathematical reasons for seeking such a 
generalization. Ferrari (1998) has noted that single--fluid models are 
unlikely to describe adequately the microphysics of extragalactic jets, and 
that a two--fluid theory would provide a useful intermediate step towards the 
development of a fully kinetic model. Moreover, the two--fluid equations 
of motion include inertial terms which, as noted above, remove troublesome 
singularities appearing in the MHD theory.       
It is convenient to have a symmetrical notation for ions (mass $m_i$,
charge $e_i=e$) and electrons (mass $m_e$, charge $e_e=-e$). In principle,
the ions can be any charged species, including positrons. Indeed, the methods
are also applicable, {\em mutatis mutandis} to quasi--neutral plasmas with 
several charged species. The equations derived below are non--relativistic, 
assume quasi-neutrality, azimuthal symmetry and steady conditions. Electron 
inertia and temperature will be included, but all dissipative and 
irreversible terms are neglected. 

As before, we use the magnetic field representation given by
equation (1). In place of the mass density $\rho$ we introduce
the common number density, $n=n_e=n_i$. We replace equation (2)
by equations for the particle flux functions, the existence of which follows 
from the continuity equations for the two species ($\chi_j \rightarrow 
\Theta_jm_j$, $j=i,e$)
$$ n {\bf v}_j = \left[ -\frac{1}{r}\frac{\partial \Theta_{j}}{\partial z} 
{\bf e}_{r}+n v_{\phi j}{\bf e}_{\phi}+\frac{1}{r}\frac{\partial \Theta_{j}}
{\partial r}{\bf e}_{z} \right], \eqno (36) $$
where $\Theta_j$, $v_{\phi j}$ are, respectively, the particle flux function 
and toroidal flow speed of species $j$. Introducing the vorticities ${\bf K}_j$
of the two species, we infer a set of relations analoguous to equations (10), 
(11) and (13):  
$${\bf K}_{j} = \left[ -\frac{\partial v_{\phi j}}{\partial z}{\bf
e}_{r}+K^{*}_{\phi j}{\bf e}_{\phi}+\frac{1}{r}\frac{\partial}{\partial r}
(rv_{\phi j}){\bf e}_{z} \right], \eqno (37) $$
$$K^{*}_{\phi j} = -\frac{1}{r}\left[\frac{\partial}{\partial
z}(\frac{1}{n}\frac{\partial \Theta_{j}}{\partial
z})+r\frac{\partial}{\partial r}(\frac{1}{rn}
\frac{\partial \Theta_{j} }{\partial r})\right], \eqno (38) $$ 
$$ {\bf K}_{j}\times n{\bf v}_{j} = (K^*_{\phi j}\frac{1}{ r}\frac{\partial
\Theta_{j}}{\partial r}-\frac{rnv_{\phi j}}{r^{2}}
\frac{\partial}{\partial r}(rv_{\phi j})){\bf e}_{r}+\frac{1}{r^{2}}
\frac{\partial (\Theta_{j},rv_{\phi j})}{\partial (r,z)}{\bf e}_{\phi}$$
$$ + (K^{*}_{\phi j}\frac{1}{r}\frac{\partial \Theta_{j}}{\partial z}-
\frac{rnv_{\phi_{j}}}{r^{2}}\frac{\partial}{\partial z}(rv_{\phi j}))
{\bf e}_{z}. \eqno (39) $$
The two equations of motion governing momentum balance now take the 
form
$$m_j{\bf K}_j\times n{\bf v}_j = -\nabla p_j-m_jn\nabla {\bf v}_j^2/2-m_jn
\nabla V-e_{j}n\nabla \Phi+e_{j}n{\bf v}_{j}\times {\bf B}/c, \eqno (40) $$
where $p_j$ is the pressure of species $j$, $e_i=e=-e_e$ and, as 
before, $\Phi$, $V$ are, respectively, electrostatic potential and 
gravitational potential. Adding the electron and ion equations of motion 
[equation (40)], we obtain the single--fluid MHD equation of motion
[equation (15)] in the formal (singular) limit $e \to \infty$ 
(keeping all other quantities fixed). This corresponds to the ion 
Larmor radius and collisionless skin depth [$c$ divided by the electron plasma
frequency $\omega_{pe}$] both tending to zero. In this limit, the electron 
equation of motion reduces to the ideal MHD Ohm's law [equation (3)].

We begin by considering the azimuthal component of equation (40). Substituting
from equation (39) and evaluating the azimuthal component of ${\bf v}_j \times
{\bf B}$, we obtain 
$$\frac{\partial (\Pi_{j},\Theta_{j})}{\partial (r,z)} = 0, \eqno (39) $$ 
where $\Pi_{j}$ are  canonical momenta of the two fluids [cf. equation (18)]:
$$\Pi_{j} = m_{j}rv_{\phi_{j}}+e_{j}\Psi/c, \eqno (42) $$ 
whence it follows that
$$\Pi_j = F_j(\Theta_j), \eqno (43) $$
where $F_j$ are arbitrary functions of the respective particle flux functions.
We now apply Amp\`ere's law, 
$$\nabla \times {\bf B} = \frac{4 \pi e}{c}(n{\bf v}_{i}-n{\bf v}_{e}). 
\eqno (44)$$
Only two of the three components of this are independent. The $r$ and $z$ 
components integrate to give
$$ rB_{\phi} = \frac{4 \pi e}{c}(\Theta_{i}-\Theta_{e}), \eqno (45) $$
while equation (8) indicates that the $\phi$ component can be written as
$$j^*_{\phi} = \frac{4 \pi en}{c}(v_{\phi i}-v_{\phi e}). \eqno (46) $$ 
Using equations (9), (42) and (43) we rewrite equation (46) in
the form
$$ \left[\frac{\partial^{2} \Psi}{\partial z^{2}}+r\frac{\partial}
{\partial r}(\frac{1}{r}
\frac{\partial\Psi}{\partial r})\right] = -\frac{4 \pi en}{c}\left[
\frac{F_{i}}{m_{i}}-\frac{F_{e}}{m_{e}}-\frac{e(m_{i}+m_{e})\Psi}
{m_{i}m_{e}c}\right]. \eqno (47) $$
This is the Grad--Shafranov equation for the two--fluid system.

We assume, in the absence of dissipation, that the entropies of the
two species are constant following the flow:
$$ \frac{\partial \sigma_{j}}{\partial l} =  0, \eqno (48) $$
where $\sigma_j=p_j/n^{\gamma}$. Taking dot products of ${\bf
v}_{j}$ with equation (40), and using equation (47), we derive two 
Bernoulli relations
$$\frac{\gamma}{\gamma -1}\frac{p_{j}}{n}+\frac{m_{j}{\bf
v}_{j}^{2}}{2}+m_{j}V+e_{j}\Phi = H_{j}(\Theta_{j}), \eqno (49) $$
where, as in the MHD case, the stagnation enthalpies $H_{j}(\Theta_{j})$ are 
arbitrarily prescribable functions of the respective particle flux functions. 
We note that these two equations effectively determine the number density $n$ 
and the electrostatic potential $\Phi$, given $\sigma_j(\Theta_{j})$, 
$F_{j}(\Theta_{j})$, $\Theta_{j}$, $\Psi$ and $V$. Since $\Phi$ 
appears nowhere else explicitly, it may be eliminated by simply adding the 
two equations to obtain the following equation for $n$:
$$ \frac{\gamma}{\gamma -1}(\sigma_{i}+\sigma_{e})n^{\gamma-1}+
(\frac{m_i{\bf v}_{i}^{2}}{2}+\frac{m_{e}{\bf
v}_e^2}{2})+(m_i+m_e)V = H_i(\Theta_i)+H_e(\Theta_e). \eqno (50) $$
Using equation (49) we may rewrite equation (40) in the form
$$(m_j{\bf K}_j+\frac{e_{j}}{c}{\bf B})\times (n{\bf v}_j) = -nH_{j}^{\prime}
\nabla \Theta_{j}+{p_j\over \gamma-1}{\sigma_j^{\prime}\over \sigma_j}\nabla
\Theta_j. \eqno (51) $$
From the radial or $z$ components of this we immediately obtain
$$ m_{j}K^{*}_{\phi j}+\frac{e_{j}}{c}B_{\phi}-\frac{n}{m_jr}F_{j}^{\prime}
(F_{j}-\frac{e_{j}}{c}\Psi) = -nrH_{j}^{\prime}+{rp_j\over \gamma-1}
{\sigma_j^{\prime}\over \sigma_j} , $$
i.e.
$$\left[\frac{\partial}{\partial z}(\frac{1}{n}\frac{\partial \Theta_{j}}
{\partial z})+r\frac{\partial}{\partial r}(\frac{1}{rn}
\frac{\partial \Theta_{j} }{\partial r})\right] -\frac{e_{j}}{m_{j}c}rB_{\phi}
+\frac{n}{m_{j}^{2}}F_{j}^{\prime}(F_{j}-\frac{e_{j}}{c}\Psi)
= nr^{2}H_{j}^{\prime}/m_{j}-{r^2n^{\gamma}\over m_j(\gamma-1)}\sigma_j^{\prime}. $$
Eliminating $rB_{\phi}$ using equation (45), we obtain the following closed 
system for $\Theta_{j}$, $\Psi$, $n$ and $V$:
$$\leftline{$\displaystyle \left[\frac{\partial}{\partial z}(\frac{1}{n}
\frac{\partial\Theta_i}{\partial z})+r\frac{\partial}{\partial r}(\frac{1}{rn}
\frac{\partial \Theta_{i} }{\partial r})\right]
-\frac{e}{m_{i}c}\frac{4 \pi e}{c}(\Theta_{i}-\Theta_{e})+\frac{n}{m_i^2}
F_i^{\prime}(F_i-\frac{e}{c}\Psi) $}$$
$$=nr^2{H_i^{\prime}\over m_i}-{r^2n^{\gamma}\over m_i(\gamma-1)}
\sigma_i^{\prime}, \eqno (52) $$
$$\leftline{$\displaystyle \left[\frac{\partial}{\partial z}(\frac{1}{n}
\frac{\partial \Theta_{e}}
{\partial z})+r\frac{\partial}{\partial r}(\frac{1}{rn}
\frac{\partial \Theta_{e} }{\partial r})\right] 
+\frac{e}{m_{e}c}\frac{4 \pi e}{c}(\Theta_{i}-\Theta_{e})
+\frac{n}{m_{e}^{2}}F_{e}^{\prime}(F_{e}+\frac{e}{c}\Psi)$}$$
$$=nr^2{H_e^{\prime}\over m_e}-{r^2n^{\gamma}\over m_e(\gamma-1)}
\sigma_e^{\prime}, \eqno (53) $$
$$\left[\frac{\partial^{2} \Psi}{\partial z^{2}}+r\frac{\partial}{\partial r}
(\frac{1}{r}\frac{\partial\Psi}{\partial r})\right] 
= -\frac{4 \pi en}{c}\left[\frac{F_{i}}{m_{i}}-\frac{F_e}{m_e}-
\frac{e(m_{i}+m_{e})\Psi}{m_{i}m_{e}c}\right], \eqno (54) $$ 
$$\frac{\gamma}{\gamma -1}(\sigma_{i}+\sigma_{e})n^{\gamma-1}+
(\frac{m_{i}{\bf v}_{i}^{2}}{2}+\frac{m_{e}{\bf v}_{e}^{2}}{2})+(m_{i}+m_{e})V
= H_{i}(\Theta_{i})+H_{e}(\Theta_{e}), \eqno (55) $$
$$\frac{1}{r}\frac{\partial}{\partial r}(r\frac{\partial V}{\partial
r})+\frac{\partial^{2} V}{\partial z^2} 
= 4\pi G\left[(m_{i}+m_{e})n+\rho_{\rm ext}\right]. \eqno (56) $$
The first two equations are the equations of motion, the third represents
Amp\`ere's law, equation (55) is 
the Bernoulli relation for the total pressure, and equation (56) is the 
gravitational Poisson equation.

Given the six arbitrary functions $F_j$, $\sigma_j$, $H_j$ and the
``external'' mass density $\rho_{\rm ext}$, these equations have to be solved 
subject to suitable boundary conditions on $\Psi$, $\Theta_j$ and the 
gravitational potential $V$. They fully determine the structure
of both flows and magnetic fields. The velocities, currents and so on 
can be obtained from the various auxiliary relations. This system represents 
an exact generalization of the equations derived by Lovelace et al. (1986) to 
two--fluid, dissipationless, azimuthally symmetric, quasi--neutral, 
nonrelativistic, gravitating equilibria with arbitrary flows. It is important 
to note that the leading order operators in the partial differential 
equations are entirely nonsingular and elliptic, provided the number density 
($n$) remains bounded. For this reason, the system is much easier to deal with 
numerically than the equations of MHD equilibrium. 

\section{Special solutions: field-aligned flows}

In this section a particular class of MHD equilibria is considered.
In Section 2 it was shown quite generally that the flow
function $\chi$ is a function of the magnetic flux, $\Psi$: we now assume that
mass density $\rho$ is also a flux function. In this case, it is easily 
shown from equations (2) and (4) that the flow velocity is divergence--free, 
i.e. the plasma is incompressible: in general, this requires that the flow 
speed $v$ be less than the sound speed $c_s$ (e.g. Landau \& Lifshitz 1987). 
Axisymmetric 
MHD equilibrium equations were obtained for general incompressible flow by 
Woltjer (1959). Here we consider incompressible flows with $\Omega \equiv 0$ 
[cf. equation (7)]: in such cases $rv_{\phi}$ is proportional to 
$rB_{\phi}$. Denoting $F^{\prime}/\rho$ by $G^{\prime}$ [cf. equations (2) and
(4)], we obtain 
$$ {\bf v} = G^{\prime}{\bf B}, \eqno (57) $$
Where $G(\Psi)$ is a new flux function. This equation indicates that the flow 
and the field are aligned in three dimensions [in general,
the alignment is in $(r,z)$ planes only]. The ideal Ohm's law [equation (3)] 
indicates that $\Phi \equiv 0$ for such flows. 

From equation (57) we infer an expression for the vorticity ${\bf K} = \nabla 
\times {\bf v}$ in terms of $G$ and {\bf B}:
$$ {\bf K} = G^{\prime \prime}\nabla \Psi \times {\bf B}+G^{\prime}
\nabla \times {\bf B}. $$ 
Substituting this in the equation of motion [equation (15)], we get
$$\rho G^{\prime}\left[G^{\prime \prime}\nabla \Psi \times {\bf B}+G^{\prime}
\nabla \times {\bf B}\right] \times {\bf B} = -\nabla p-\rho \nabla 
\frac{{\bf v}^{2}}{2}-\rho \nabla V+\frac{1}{4\pi} (\nabla \times {\bf B})
\times {\bf B}. $$
Using ${\bf B}\cdot\nabla\Psi=0$, we find that the above equation may
be written in the form
$$-\rho G^{\prime}G^{\prime \prime} {\bf B}^{2} \nabla \Psi +(\rho G^{\prime
2}-\frac{1}{4\pi})(\nabla \times {\bf B})\times {\bf B}
= -\nabla p-\rho \nabla \frac{{\bf v}^{2}}{2}-\rho \nabla V. \eqno (58) $$
If we take the dot product of this equation with ${\bf B}$, consistency 
requires that the right hand side must vanish. Using the fact that $\rho$ is a
flux function, we infer from this that  
$${\bf B}\cdot\nabla (p+\rho  \frac{{\bf v}^{2}}{2}+\rho V) = 0. $$
We thus obtain the Bernoulli relation
$$p+\rho \frac{{\bf v}^{2}}{2}+\rho V = H^*(\Psi), \eqno (59) $$
where $H^*$ is an arbitrary function of $\Psi$. Using this to eliminate $p$ 
from equation (58), we obtain
$$-\rho G^{\prime}G^{\prime \prime} {\bf B}^2\nabla\Psi+(\rho G^{\prime
2}-\frac{1}{4\pi})(\nabla \times {\bf B})\times {\bf B}
= -H^{* \prime}\nabla \Psi +\rho^{\prime}\nabla \Psi (\frac{{\bf v}^{2}}{2}+ 
V). $$
Putting ${\bf v} = G^{\prime}{\bf B}$ [equation (57)], this reduces to 
$$-(\rho G^{\prime}G^{\prime \prime}+\rho^{\prime}\frac{G^{\prime 2}}{2})
{\bf B}^{2}\nabla \Psi +(\rho G^{\prime 2}-\frac{1}{4\pi})(\nabla \times 
{\bf B})\times {\bf B} = -(H^{* \prime}- \rho^{\prime} V) \nabla \Psi.
\eqno (60) $$ 
This is our fundamental equation. We first note that since $\Psi$ depends only
on $r$ and $z$, the azimuthal component is satisfied if 
${\bf j}\times {\bf B}$ vanishes in the $\phi$ direction. 
It is clear from equation (12) that this requires $rB_{\phi}$ to be a flux 
function, which we identify with $I(\Psi)$ defined by equation (23) [$\Theta 
\equiv 0$ in this case]. It follows from our field alignment assumption that 
$rv_{\phi}=J(\Psi)=G^{\prime}I$. We observe next that if $\rho
G^{\prime 2}$ is a constant, the first term on the left hand side 
of equation (58) vanishes. In this case, making use of 
the $r$ or $z$ components of equation (12), we find that the equation reduces 
to
$$\lambda (j^{*}_{\phi}\frac{1}{r}-\frac{I I^{\prime}}{r^{2}})
= -(H^{* \prime}- \rho^{\prime} V), \eqno (61) $$  
where
$$\lambda = (\rho G^{\prime 2}-\frac{1}{4\pi}). $$
Equation (61) can be simplified, using equation (9), to give a Grad--Shafranov
equation for field--aligned MHD flows:
$$\lambda \left[\frac{\partial^{2} \Psi}{\partial z^{2}}+r\frac{\partial}
{\partial r}(\frac{1}{r}\frac{\partial
\Psi}{\partial r})+I I^{\prime}\right]+r^{2} \rho^{\prime} V
= r^{2}H^{* \prime}. \eqno (62) $$
This is algebraically similar to the Grad--Shafranov equation used by 
Bogoyavlenskiy (2000), but represents a generalization of it to include
field--aligned flow in a spatially--varying gravitational potential. From the 
definition of $G^{\prime}$ [equation (57)], it is clear that 
$4\pi\rho G^{\prime 2} = v^2/c_A^2$, where $c_A = B/\sqrt{4\pi\rho}$ is the 
Alfv\'en speed. Although we assumed that $\rho G^{\prime 2}$ was a constant in
order to obtain equation (62), this does not, of course, preclude the 
possibility of $v$, $B$ and $\rho$ individually varying in space.
   
There are several limiting cases of equation (62) which are of physical 
interest. If $V$ is a prescribed gravitational potential, $\rho$ and $H^{*}$ 
are quadratic in $\Psi$, and $II^{\prime}$ is linear in $\Psi$, we obtain 
a linear eigenvalue problem for $\Psi$ and $\lambda$,
which can be solved by a variety of methods (e.g. Bogoyavlenskij 2000). Even 
when equation (60) is nonlinear, it can be solved numerically in a 
straightforward way. 

A special case of great interest is that of constant
density and force--free (Beltrami) magnetic fields, defined by the 
property $\nabla\times {\bf B} = \mu {\bf B}$ for some scalar function $\mu$,
so that $(\nabla\times {\bf B})\times {\bf B} = {\bf 0}$ (Lundquist 1951;
Mahajan \& Yoshida 1998). 
Consistency with equation (60), with constant $\rho G^{\prime 2} \ne 1/4\pi$, 
requires that 
$H^*$ also be a constant. More generally, the force--free condition is 
satisfied for constant $\rho G^{\prime 2}$ whenever $H^{* \prime}=
\rho^{\prime}V$. From equation (62), it is clear that $\lambda$ is then an 
arbitrary finite constant, and that $\Psi$ satisfies the equation
$$ \frac{\partial^{2} \Psi}{\partial z^{2}}+r\frac{\partial}{\partial r}
(\frac{1}{r}\frac{\partial\Psi}{\partial r})+II^{\prime} = 0. \eqno (63)$$
Having solved equation (63) for $\Psi$, one can obtain the velocity field
{\bf v}$(r,z)$ from equation (57): the pressure is then determined by the 
Bernoulli relation [equation (59)], as befits an incompressible flow.

A second class of solutions has $\lambda=0$. The equilibrium is then
compatible with a completely arbitrary azimuthally symmetric field! 
It is clear from equation (60) that it is not necessary for the azimuthal
component of ${\bf j}\times {\bf B}$ to vanish in this case: the only 
consistency requirements are that $\rho G^{\prime 2}$ is a constant and that 
$H^{* \prime}=\rho^{\prime}V$. Since $\lambda = \rho G^{\prime 2}-1/4\pi = 
(v^2/c_A^2-1)/4\pi$, the condition $\lambda = 0$ means that the flow is 
everywhere ``trans--Alfv\'enic'' ($v = c_A$), as well as being field--aligned. 
This solution was first obtained by Chandrasekhar (1956), who proved moreover 
that it is stable. Consistency with the assumption of incompressibility 
($\nabla\cdot{\bf v}=0$) requires in this case that $c_s > c_A$, i.e. that the
plasma beta $\beta \sim c_s^2/c_A^2$ be greater than unity. It is not clear 
how restrictive this condition is, since $\beta$ in most astrophysical 
plasmas is not accurately known.     

The most general field--aligned flow governed by equation (59) is one in which
$\lambda$, and hence $\rho G^{\prime 2}$, are flux functions, the equation 
then taking the form
$$\lambda(\Psi) \left[\frac{\partial^{2} \Psi}{\partial z^{2}}+r\frac{\partial}
{\partial r}(\frac{1}{r}\frac{\partial\Psi}{\partial r})+II^{\prime}\right]+
\frac{1}{2}\lambda^{\prime}((\nabla \Psi)^{2}+I^{2}) = r^{2}(H^{* \prime}-
\rho^{\prime}V), \eqno (64)$$ 
where
$$\lambda(\Psi) = \rho G^{\prime 2}-\frac{1}{4\pi}. \eqno (65) $$
It is plain that we can extend the same ideas to the two--fluid case by taking
$n$ to be a flux function and ${\bf v}_j=G^{\prime}_j(\Psi){\bf B}$.
Specific applications of these solutions are left to future work.

\section{Conclusions and Discussion}

We have derived equilibrium equations for azimuthally symmetric MHD and 
two--fluid systems with arbitrary nonrelativistic flows. Our method of 
derivation in the MHD case is distinct from that used by Lovelace et al. 
(1986), and provides the basis for our two--fluid calculation.
We have identified some special exact solutions of the MHD system of 
equations which are of independent interest. Since there are no restrictions 
on the geometry (other than azimuthal symmetry), both systems of equations are
equally applicable to astrophysical and laboratory systems with azimuthal 
symmetry. The two--fluid model is more suitable than the MHD model for 
astrophysical applications, not only because it contains more physics, but 
also from the computational point of view: the two--fluid equations do not 
involve singularities which are associated with ideal MHD (Blandford \& Payne 
1982; Lovelace et al. 
1986). The physical reason for this is that the
two--fluid model, unlike the MHD model, takes into account electron 
inertia, which introduces a new fundamental length, the collisionless skin
depth $c/\omega_{pe}$ (in the MHD limit, $c/\omega_{pe} \to 0$).  
In future work we intend to solve numerically both the MHD and 
two--fluid systems of equations for specific astrophysical scenarios,
including extragalactic jets. 

The equations we have derived are nonrelativistic. Flow speeds much less than 
$c$ occur, for example, in jets, hot spots and lobes associated with radio 
galaxies, but close to the core $v \sim c$ (Ferrari 1998). If the assumptions 
of stationarity and azimuthal symmetry are retained, it is 
reasonable to expect that similar but more complicated equations will exist 
for relativistic conditions (this was demonstrated by Lovelace et al. for the 
case of ideal MHD). Although we considered plasmas with only one or two 
separate species, the methods clearly extend to any number of charged species,
provided that overall quasi--neutrality and other equilibrium assumptions 
apply. It is relatively straightforward to take into account higher order 
collisional or collisionless effects (using, for example, the reciprocals of 
the Reynolds or Lundquist numbers as expansion parameters): one would expect
the inclusion of such effects to provide transport equations for the various 
arbitrary functions appearing in the MHD and two--fluid systems of equations.  

\section*{Acknowledgements}

We are grateful to Drs R. O. Dendy, C. M. Roach, M. Tagger and A. Webster for 
helpful discussions. This work was supported in part by the Commission of the 
European Communities under TMR Network Contract ERB--CHRXCT980168, by the UK 
Department of Trade and Industry, and by EURATOM.

\section*{References}

\ref{Blandford R. D., Payne D. G., 1982, MNRAS, 199, 883}

\ref{Bogoyavlenskij O. I., 2000, Phys. Rev. Lett., 84, 1914}

\ref{Camenzind M., 1998, in Massaglia S., Bodo G., eds., Astrophysical 
Jets: Open Problems. Gordon and Breach, Amsterdam, p. 3}

\ref{Chandrasekhar S., 1956, ApJ, 124, 232}

\ref{Ferrari A., 1998, ARA\&A, 36, 539} 

\ref{Freidberg J. P., 1982, Rev. Mod. Phys., 54, 801}

\ref{Goedbloed J. P., Lifschitz A., 1997, Phys. Plasmas, 4, 3544}

\ref{Grad H., Rubin H., 1958, Proceedings of the second United Nations 
International Conference on the Peaceful Uses of Atomic Energy, Vol. 31. 
United Nations, Geneva, p. 190}

\ref{Hawley J.F., Stone J. M., 1998, ApJ, 501, 758}

\ref{Henriksen R. N., 1998, in Massaglia S., Bodo G., eds., Astrophysical 
Jets: Open Problems. Gordon and Breach, Amsterdam, p. 191}

\ref{Kellet B. J., Bingham R., Tsikoudi V., 2000, MNRAS, in press} 

\ref{Keppens R., Goedbloed J. P., 2000, ApJ, 530, 1036}

\ref{Krasheninnikov S. I., Catto P. J., Hazeltine R. D., 2000, Phys. Plasmas,
7, 1452}

\ref{Landau L. D., Lifshitz E. M., 1987, Fluid Mechanics, 2nd Edition.
Pergamon, Oxford, p. 21}  

\ref{Lovelace R. V. E., Mehanian C., Mobarry C. M., Sulkanen M. E., 1986, 
ApJS, 62, 1}

\ref{Maschke E. K., Perrin, 1980, Plasma Phys., 22, 579}

\ref{Lundquist S., 1951, Phys. Rev. 83, 307}

\ref{Mahajan S. M., Yoshida Z., 1998, Phys. Rev. Lett. 81, 4863} 

\ref{Meyer R. E., 1971, Introduction to Mathematical Fluid Dynamics. 
Wiley--Interscience, London, p. 157}

\ref{Morozov A. I., Solov'ev L. S., 1963, Sov. Phys. Doklady, 8, 243} 

\ref{Rosso F., Pelletier G., 1994, A\&A, 287, 325}

\ref{Shafranov V.D., 1958, Sov. Phys. JETP 6, 545}

\ref{Tagger M., Pellat R., 1999, A\&A 349, 1003}

\ref{Throumoulopoulos G. N., Pantis G., 1989, Phys. Fluids B 1, 1827}

\ref{Wesson J., 1997, Tokamaks, 2nd Edition. Oxford University Press, Oxford}

\ref{Woltjer L., 1959, ApJ, 130, 400}

\ref{Zehrfeld H. P., Green B. J., 1972, Nucl. Fusion, 12, 569}

\ref{Zensus J. A., 1997, ARA\&A, 35, 607}

\ref{Zweibel E. G., Heiles C., 1997, Nature, 385, 131}

\end{document}